\documentclass[rotating,dvips]{arxstspdf}
\usepackage{dcolumn}
\usepackage{flushend}

\volume{23}
\issue{3}
\pubyear{2008}
\firstpage{313}
\lastpage{317}
\doi{10.1214/08-STS244B}
\referstodoi{10.1214/07-STS244}

\makeatletter
\newcommand{\ldel}{\stackrel{\mbox{\scriptsize{11}}}{\ldots}}
\newcommand{\ldtw}{\stackrel{\mbox{\scriptsize{12}}}{\ldots}}
\def\eqref#1{(\ref{#1})}
\newcolumntype{d}[1]{D{.}{.}{#1}}
\makeatother

\begin{document}
\begin{frontmatter}

\title{Comment: Quantifying Information Loss in Survival Studies}
\runtitle{Comment}

\begin{aug}
\author{\fnms{Hani} \snm{Doss}\ead[label=e1]{doss@stat.ufl.edu}}
\runauthor{H. Doss}

\affiliation{University of Florida}

\address{Hani Doss is Professor, Department of
Statistics, University of Florida, Gainesville, Florida 32611, USA \printead{e1}.}

\end{aug}

% ABSTRACT

% KEYWORDS

\end{frontmatter}

In their paper, Nicolae, Meng and Kong (henceforth NMK) propose
several very interesting methods for quantifying the fraction of
missing information in a sample, and focus their attention on
genetic studies.  Survival analysis is another area in statistics
where missing information plays an important role.  Here, censoring
complicates study design, for example when we want to determine how
big a clinical trial should be in order to have a good chance of
detecting a treatment effect in a Cox model.  Most current methods
for dealing with this difficult problem involve two stages, where in
the first stage we make a projection of what the variance of the
coefficient of the treatment effect would be if there was no
censoring, and in the second stage we make a correction to adjust
for the censoring.  Often this is done under restrictive parametric
(e.g., exponential) assumptions for the underlying distributions.
It would be desirable to use the methods proposed by NMK in the
survival analysis setting.  I tried to carry over their methods to
the Cox model, and encountered some problems.  The difficulties I
discovered led me to consider modifications of their proposals,
which I believe work well.  Below I discuss the setup I consider, my
experiences, the issues, and some approaches I think are promising.

%s1 ###
\section{Survival Studies for Assessing the Efficacy of a New
Treatment}
\label{sec:intro}
A typical clinical trial with a survival outcome involves a fixed
time frame, say five years.  Patients enter the trial continuously
during the first four years, are randomly assigned to treatment or
control, and the last year is a followup year, during which no
patients enter the study.  Some patients die during the study, in
which case their survival time is observed.  But some patients die
from other causes or are lost to followup, and some are still alive
at the time the trial is ended; so in these cases the survival time
is censored: for each individual in this group, there is a time $t$
and we know only that the individual's survival is greater than $t$.

Clearly the censoring reduces information regarding the efficacy of
the new treatment.  When designing a subsequent study in the hope of
getting stronger evidence against the null hypothesis of no
treatment effect, we now have two choices: increase the number of
patients in the study, which can be expensive, or try to reduce the
censoring.  We can reduce the censoring either by putting more
resources into followup, or by extending the length of the period of
time after the end of the accrual period.  These result in costs
which are financial and also ethical because increasing the length
of the final followup period postpones publication of results that
are of potential benefit to other patients.  The decision of whether
to increase the number of patients or to reduce the censoring
depends crucially on the amount of information loss due to
censoring, so being able to measure this is extremely important in
the design of future studies.  This situation is very similar to the
one discussed by NMK.

By far the most commonly used model for regression with censored
survival data is the Cox proportional hazards model.  Suppose that
individual $i$ has covariate~vector $Z_i = (Z_{i1},\ldots,
Z_{ip})$, where $Z_{i1}$ is the indicator that the individual
receives the treatment.  Let $X_i$ be the death time of individual
$i$ if there was no censoring, and let
 $Y_i$ be the censoring time.
For each individual, we observe the minimum $T_i = \min(X_i, Y_i)$
and also the indicator $\delta_i$ that $X_i$ was not censored, that is,
$\delta_i = I(X_i \leq Y_i)$.  So the data for individual $i$ is the
triple $(T_i, \delta_i, Z_i)$.

The proportional hazards model stipulates that the hazard rate for
an individual with covariate vector $Z$ is given by
%
%e1 ###
\begin{equation}
  \label{cox-model-1}
  \lambda(t|Z) = \lambda_0(t) \exp(\beta' Z),
\end{equation}
where $\beta$ is a $p$-dimensional vector of coefficients, and
$\lambda_0$ is the hazard function for an individual with covariate
vector $0$.  For our purposes (as will be clear later), it is
preferable to define the model in terms of cumulative hazard
functions, and so by integrating~(\ref{cox-model-1}), the model is
stated by specifying that $\Lambda(t|Z)$, the cumulative
hazard function for an individual with covariate $Z$, is related to
$\Lambda_0(t)$, the cumulative hazard function for an individual
with covariate $0$ via
%
%e2 ###
\begin{equation}
  \label{cox-model-2}
  \Lambda(t |  Z) = \Lambda_0(t) \exp(\beta'Z).
\end{equation}
The model is parameterized by $\theta = (\Lambda_0, \beta)$, in
which $\Lambda_0$ is considered a nuisance parameter.  The
likelihood function is very complex, and involves both $\Lambda_0$
and~$\beta$.  Cox's partial likelihood (Cox, \citeyear{Cox1972}, \citeyear{Cox1975})---literally just a part of
the full likelihood; see Efron (\citeyear{Efron1977})---involves only $\beta$.

%s2 ###
\section{Measuring the Relative Information in the Data}
\label{measuring-info}
There is a large literature that shows that Cox's partial likelihood
has the main features of an ordinary likelihood: the maximum partial
likelihood estimator~$\hat{\beta}$ is consistent and asymptotically
normal (Andersen and Gill, \citeyear{AndersenGill1982}), and there are several papers
(Efron, \citeyear{Efron1977}; Oakes, \citeyear{Oakes1977}) that show that
inference based on this partial likelihood is essentially as good as
inference based on the full likelihood.  Standard software gives the
partial likelihood function.  For example in R, if we fit a Cox
model to a data set and call the result \texttt{fitcox}, then
\texttt{fitcox\$loglik} gives the log of the partial likelihood,
evaluated at any desired value of $\beta$, and also at the maximum
partial likelihood estimate of~$\beta$.

These considerations suggest that we use the partial likelihood
function as a likelihood in the measure $\mathcal{R}I_1$ that NMK
propose.  For a data set $D$, let $\ell_D(\beta)$ denote the log
partial likelihood function based on $D$.  Let $D_{\mathrm{ob}}$
denote the observed data, and $D_{\mathrm{co}}$ denote the full data,
had we been able to see it.  Suppose we wish to test the null
hypothesis that $\beta = \beta_0$.  If we use the partial
likelihood, the numerator of $\mathcal{R}I_1$ is simply
$\ell_{D_{\mathrm{ob}}}(\hat{\beta}) - \ell_{D_{\mathrm{ob}}}(\beta_0)$,
and the denominator is
%
%e3 ###
\begin{equation}
  \label{RI1-den}
  E_{\hat{\theta}} \bigl\{ \ell_{D_{\mathrm{co}}}(\hat{\beta}) \, | \,
  D_{\mathrm{ob}} \bigr\} - E_{\hat{\theta}} \bigl\{
  \ell_{D_{\mathrm{co}}}(\beta_0) \, | \, D_{\mathrm{ob}} \bigr\}.
\end{equation}
In~\eqref{RI1-den}, $D_{\mathrm{co}}$ is random and has the
conditional distribution of the complete data given the observed
data, and the subscript $\hat{\theta}$ indicates that this
conditional distribution is computed under the assumption that
$\hat{\theta}$ is the true value of $\theta$.  Here, the maximum
likelihood estimator of~$\theta$ is $\hat{\theta} =
(\hat{\Lambda}_0, \hat{\beta})$, where $\hat{\Lambda}_0$ is the
Nelson--Aalen estimator of $\Lambda_0$.  This expectation is
hopelessly difficult to compute.  However, it is possible to
estimate it via Monte Carlo, and the last section of this article
details how to do this.

To assess the performance of this measure I considered the ``acute
myelogenous leukemia data'' and some perturbations of it.  This data
set is given in Miller (\citeyear{Miller1981}, page 49), and is available in the
\texttt{survival} package in R\@.  There are $11$ individuals
receiving the new treatment $(Z = 0)$, of whom four have censored
survival times, and $12$ individuals receiving the standard
treatment $(Z = 1)$, of whom one has a censored survival time.  We
are interested in testing the null hypothesis that $\beta = 0$,
indicating no treatment effect.

%t1 ###
\begin{table*}%[h]
\tabcolsep=0pt
\caption{Three versions of the leukemia data}\label{tab:aml-data}%@{\hspace{2mm}}p{4mm}
\begin{tabular*}{177mm}{@{\extracolsep{\fill}}lcccccd{2.-1}d{2.-1}d{2.-1}d{2.-1}d{2.-1}d{2.-1}d{2.-1}d{2.-1}d{2.-1}d{3.-1}cccccccd{2.-1}d{2.-1}d{2.-1}d{2.-1}d{2.-1}d{2.-1}d{2.-1}d{2.-1}@{}}
\hline
                        & $T$      &   &         &   & 9 & 13 & 13 & 18 & 23 & 28 & 31 & 34 & 45 & 48 & 161 &   &         &   & 5 & 5 & 8 & 8 & 12 & 16 & 23 & 27 & 30 & 33 & 43 & 45 \\
  \textsf{aml-orig} & $\delta$ &   &         &   & 1 & 1  & 0  & 1  & 1  & 0  & 1  & 1  & 0  & 1  & 0   &   &         &   & 1 & 1 & 1 & 1 & 1  & 0  & 1  & 1  & 1  & 1  & 1  & 1  \\
                        & $Z$      &   &         &   & 0 & 0  & 0  & 0  & 0  & 0  & 0  & 0  & 0  & 0  & 0   &   &         &   & 1 & 1 & 1 & 1 & 1  & 1  & 1  & 1  & 1  & 1  & 1  & 1  \\ [4mm]
                      & $T$      & 0 & $\ldel$ & 0 & 9 & 13 & 13 & 18 & 23 & 28 & 31 & 34 & 45 & 48 & 161 & 0 & $\ldtw$ & 0 & 5 & 5 & 8 & 8 & 12 & 16 & 23 & 27 & 30 & 33 & 43 & 45 \\
  \textsf{aml-1}    & $\delta$ & 0 & $\ldel$ & 0 & 1 & 1  & 1  & 1  & 1  & 1  & 1  & 1  & 1  & 1  & 1   & 0 & $\ldtw$ & 0 & 1 & 1 & 1 & 1 & 1  & 1  & 1  & 1  & 1  & 1  & 1  & 1  \\
                        & $Z$      & 0 & $\ldel$ & 0 & 0 & 0  & 0  & 0  & 0  & 0  & 0  & 0  & 0  & 0  & 0   & 1 & $\ldtw$ & 1 & 1 & 1 & 1 & 1 & 1  & 1  & 1  & 1  & 1  & 1  & 1  & 1  \\ [4mm]
                        & $T$      & 0 & $\ldel$ & 0 & 9 & 13 & 13 & 18 & 23 & 28 & 31 & 34 & 45 & 48 & 161 & 0 & $\ldtw$ & 0 & 5 & 5 & 8 & 8 & 12 & 16 & 23 & 27 & 30 & 33 & 43 & 45 \\
  \textsf{aml-2}    & $\delta$ & 0 & $\ldel$ & 0 & 1 & 1  & 0  & 1  & 1  & 0  & 1  & 1  & 0  & 1  & 0   & 0 & $\ldtw$ & 0 & 1 & 1 & 1 & 1 & 1  & 0  & 1  & 1  & 1  & 1  & 1  & 1  \\
                        & $Z$      & 0 & $\ldel$ & 0 & 0 & 0  & 0  & 0  & 0  & 0  & 0  & 0  & 0  & 0  & 0   & 1 & $\ldtw$ & 1 & 1 & 1 & 1 & 1 & 1  & 1  & 1  & 1  & 1  & 1  & 1  & 1  \\
\hline
  \end{tabular*}
\tabnotetext[]{tz}{Notation of the
 sort $0$ $\ldel$
  $0$ indicates a string of $11$ $0$'s.}
\end{table*}

Table~\ref{tab:aml-data} gives three versions of this data
set, of which the first is the original data set.  Dataset
\textsf{aml-1} is a perturbed version in which (i) all the status
indicators $\delta_i$ that were $0$ were changed to $1$ and (ii)
$11$ observations, all censored at time $0$, were added to the new
treatment group, and $12$ observations, all censored at time $0$,
were added to the standard treatment group.  The inclusion of these
$23$ new observations all censored at time $0$ doubles the size of
the data set but adds no information whatsoever, and any reasonable
method for estimating the relative information in the data should
give $0.5$.  This is the censored data analogue of the example of
unobserved Bernoullis in Section~1.3 of NMK\@.  Dataset
\textsf{aml-2} is a perturbed version of the original data set in
which $11$ observations, all censored at time $0$, are added to the
new treatment group, and $12$ observations, all censored at time
$0$, are added to the standard treatment group; but the original
part of the data set was not altered.

The results are given in line~$1$ of Table~\ref{tab:results}.
They are surprising.  The value of $\mathcal{R}I_1$ for the original
data set is $0.987$, suggesting that there is essentially no missing
information, even though $5$ of the $23$ observations are censored;
and for \textsf{aml-1}, the value is $0.552$ whereas it should be
$0.5$, or at least very close to $0.5$; and what is more worrisome is
that for \textsf{aml-2} it is bigger than for \textsf{aml-1}, even
though \textsf{aml-2} has more missing data.  In fact, it is not
even true that $\mathcal{R}I_1$ is always less than $1$. (A particular
instance of this phenomenon arises when dealing with the data set
\textsf{veteran}, available in the survival package in R, when
testing whether the treatment effect is $0$, and ignoring all other
predictors.)

An explanation for this is as follows.  The partial likelihood uses
only the information at the times of the uncensored deaths
(Efron, \citeyear{Efron1977}), whereas the full likelihood also includes the
information between successive uncensored deaths.  The data used to
form the denominator of $\mathcal{R}I_1$ involves some censored
observations, whereas the data used to form the numerator does not.
So it appears that the parts missing from the partial likelihood are
different in the numerator and denominator of $\mathcal{R}I_1$.  This
is a very rare instance where using the partial likelihood creates
serious problems.  The net effect is that the key inequality (16) in
NMK fails: the inequality is based on using the full likelihood.
Consequently the basic inequality $\mathcal{R}I_1 \leq 1$ need not
hold.

The rationale for the criterion $\mathcal{R}I_1$ suggests the following
alternative way of forming the ratio of\break ``evidence against the null
hypothesis in the present sample'' to the ``expected value of the
evidence against the null hypothesis if we had the complete data
set,'' which bypasses the likelihood function.  For a given method
of estimating $\theta$ and a data set $D$, let\, $\hat{\theta}(D)$
denote the estimate based on data $D$, and  let$\ \widehat{\operatorname{Var}}
(\hat{\theta}(D))$ be an estimate of the covariance matrix of
$\hat{\theta}(D)$.  Also let $V_{\mathrm{ob}} = \widehat{\operatorname{Var}}
(\hat{\theta} (D_{\mathrm{ob}}))$ and $V_{\mathrm{co}}$ be the matrix
whose inverse is given by
%
%e4 ###
\begin{equation}
  \label{vars-ob-co}
  V_{\mathrm{co}}^{-1} = E_{\hat{\theta}} \{
  [\widehat{\operatorname{Var}} (\hat{\theta}
  (D_{\mathrm{co}}))]^{-1} | D_{\mathrm{ob}} \},
\end{equation}
where, as before, $D_{\mathrm{ob}}$ is the complete data, and
$D_{\mathrm{co}}$ is random and has the conditional distribution of
the complete data given the observed data; and the subscript
$\hat{\theta}$ indicates that this conditional distribution is
computed under the assumption that $\hat{\theta}$ is the true value
of $\theta$.  We form
%
%e5 ###
\begin{equation}
  \label{RIW}
  \mathcal{R}I_W = \frac { ( \hat{\theta} (D_{\mathrm{ob}}) -
  \theta_0 )' \, V_{\mathrm{ob}}^{-1} \, ( \hat{\theta}
  (D_{\mathrm{ob}}) - \theta_0 ) } { ( \hat{\theta}
  (D_{\mathrm{ob}}) - \theta_0 )' \, V_{\mathrm{co}}^{-1} \, (
  \hat{\theta} (D_{\mathrm{ob}}) - \theta_0 ) },
\end{equation}
which is a ratio of Wald-like quantities.  If the dimension of
$\theta$ is $1$, the reciprocal of $\mathcal{R}I_W$ simplifies to
\[
  (\mathcal{R}I_W)^{-1} = \frac{V_{\mathrm{ob}}}{V_{\mathrm{co}} } =
  E_{\hat{\theta}} \biggl\{ \frac { \widehat{\operatorname{Var}}
  (\hat{\theta} (D_{\mathrm{ob}})) } { \widehat{\operatorname{Var}}
  (\hat{\theta} (D_{\mathrm{co}})) } \, \bigg | \, D_{\mathrm{ob}}
  \biggr\}
\]
and has the interpretation of ``expected value of the ratio of the
variance of the $\hat{\theta}$ we have to the variance of what
$\hat{\theta}$ would be if we had the complete data.''  Motivation
for~\eqref{RIW} in general is given at the end of this section.

%t2 ###
\begin{table*}[t]
\caption{Monte Carlo estimates of the $\mathcal{R}I_1$ and $\mathcal{  R}I_W$ criteria,
together with $99\%$ confidence intervals, on
  three versions of the leukemia data}
  \label{tab:results}
    \begin{tabular*}{\textwidth}{@{\extracolsep{\fill}}lccc@{}}
    \hline
 & \textsf{aml-orig} & \textsf{aml-1}  & \textsf{aml-2}  \\
  \hline
  $\mathcal{R}I_1$ & $0.987$ $(0.976, 0.999)$ & $0.552$ $(0.538, 0.567)$ & $0.694$ $(0.675, 0.714)$ \\
  $\mathcal{R}I_W$ & $0.847$ $(0.844, 0.849)$ & $0.490$ $(0.489, 0.491)$ & $0.387$ $(0.386, 0.389)$\\
  \hline
  \end{tabular*}
\tabnotetext[]{tz}{Each case is obtained by a
  Monte Carlo run of $5000$ simulations, as described in
  Section~\ref{gen-cd}, and takes about a minute to produce on a
  $3.8$-GHz dual core P$4$ running Linux.  For $\mathcal{R}I_W$, the
  estimates are very stable: $99\%$ confidence intervals have width
  of about $0.003$.  For $\mathcal{R}I_1$, the confidence intervals are
  wider.}
\end{table*}

When we apply this criterion to the example of unobserved Bernoullis
in Section~1.3 of NMK, a short calculation shows that this approach
gives what $\mathcal{R}I_1$ gives, namely that the fraction of
information in the sample is $n_0/n$ (to order $1/n$).

Line~$2$ of Table~\ref{tab:results} gives the value of $\mathcal{R}I_W$
for the three versions of the leukemia data, when we estimate~$\beta$ via the maximum partial likelihood estimator, and the
variance estimate is the negative second derivative of the log
partial likelihood function at its maximum.  The pattern we see
makes sense.  For \textsf{aml-orig}, which includes five partially
informative censored observations, $\mathcal{R}I_W$ gives a number
intermediate between $1$ and the proportion of uncensored
observations ($0.783$); it is almost equal to $0.5$ for
\textsf{aml-1}, correctly reflecting the fact that the additional
$23$ points censored at $0$ give no information at all; and it is
less than $0.5$ for \textsf{aml-2}, which includes not only $23$
completely uninformative points, but also the original censored
observations.  It should be noted that the variances used in the
calculation of $V_{\mathrm{ob}}$ and $V_{\mathrm{co}}$ are estimated
variances, and so the value of $\mathcal{R}I_W$ depends on the
particular estimate that is used.  This dependence may be noticeable
in small samples.  For instance, this is the reason why $\mathcal{R}I_W$ gives $0.490$ instead of $0.5$ for \textsf{aml-1}.
Table~\ref{tab:results} gives results for a single experiment, but I
got very similar results for many other data sets, including data
sets that are bigger, have a bigger percentage of censored
observations, or both.

Criterion~\eqref{RIW} has the following advantages:

\begin{itemize}
\item[$\bullet$] It does not require the evaluation of a likelihood at some
  estimate.  In fact, $\hat{\theta}$ need not be a maximum
  likelihood estimator, and there need not even be a likelihood
  function.  This is important for some situations---for example
  when we have a single randomly censored sample and we use the
  Kaplan--Meier estimate---when there is no likelihood at all.
\item[$\bullet$] It handles nuisance parameters without modification.  That is,
  if $\theta = (\theta^{(1)}, \theta^{(2)})$, and the null
  hypothesis involves only $\theta^{(1)}$, then we simply
  form~\eqref{RIW} with~$\theta^{(1)}$ and $\theta^{(1)}_0$
  replacing $\hat{\theta}$ and $\theta_0$, etc.
\end{itemize}

To motivate~\eqref{RIW}, suppose we are in a parametric framework,
and recall that $\mathcal{R}I_1$ is given by
%
%e6 ###
\begin{equation}
  \label{RI1}
  \frac
  {  \ell_{D_{\mathrm{ob}}} (\hat{\theta} (D_{\mathrm{ob}})) -
  \ell_{D_{\mathrm{ob}}} (\theta_0)  }
  {  E_{\hat{\theta}} \{ \ell_{D_{\mathrm{co}}} (\hat{\theta}
  (D_{\mathrm{ob}})) - \ell_{D_{\mathrm{co}}} (\theta_0) |
  D_{\mathrm{ob}} \}  },
\end{equation}
and let us compare this to the closely related quantity
\begin{eqnarray}
  \label{RIW-alt}
  &&\mathcal{R}I_{W\mbox{-}\mathrm{alt}}\nonumber\\
   &&\quad {}= \bigl( \hat{\theta} (D_{\mathrm{ob}}) - \theta_0 \bigr)' [
  -\ddot{\ell}_{D_{\mathrm{ob}}} (\hat{\theta} (D_{\mathrm{ob}}))]\bigl( \hat{\theta}(D_{\mathrm{ob}}) - \theta_0 \bigr)
  \nonumber\\[-6pt]\\[-6pt]
  &&\qquad {}/\bigl(E_{\hat{\theta}} \bigl\{ \bigl( \hat{\theta} (D_{\mathrm{ob}}) -
  \theta_0 \bigr)'[ -\ddot{\ell}_{D_{\mathrm{co}}}
  (\hat{\theta} (D_{\mathrm{co}}))]\nonumber\\
  &&\quad \hspace*{98pt}\cdot\bigl(
  \hat{\theta}(D_{\mathrm{ob}}) - \theta_0 \bigr) \big|
  D_{\mathrm{ob}} \bigr\}\bigr),\nonumber
\end{eqnarray}
in which $\ddot{\ell}_{D_{\mathrm{ob}}}$ denotes the second derivative
(with\break respect to $\theta$) of $\ell_{D_{\mathrm{ob}}}$.  Consider the
numerator of~\eqref{RI1}.  Assuming standard regularity conditions,
a two-term Taylor expansion of $\ell_{D_{\mathrm{ob}}}(\theta_0)$
around $\hat{\theta} (D_{\mathrm{ob}})$ gives the numerator
of~\eqref{RIW-alt} (except for a factor of $2$).  If we expand
$\ell_{D_{\mathrm{co}}} (\theta_0)$ around $\hat{\theta}
(D_{\mathrm{ob}})$ and approximate $\dot{\ell}_{D_{\mathrm{co}}}
(\hat{\theta} (D_{\mathrm{ob}}))$ and $\ddot{\ell}_{D_{\mathrm{co}}}
(\hat{\theta} (D_{\mathrm{ob}}))$ by $\dot{\ell}_{D_{\mathrm{co}}}
(\hat{\theta} (D_{\mathrm{co}}))$ and $\ddot{\ell}_{D_{\mathrm{co}}}
(\hat{\theta} (D_{\mathrm{co}}))$, respectively, the denominator
of~\eqref{RI1} is the denominator of~\eqref{RIW-alt} (except for a
factor of $2$), and in~\eqref{RIW-alt} we may take $(
\hat{\theta} (D_{\mathrm{ob}}) - \theta_0 )'$ and $(
\hat{\theta} (D_{\mathrm{ob}}) - \theta_0 )$ outside the
expectation.  Expressions~\eqref{RIW} and~\eqref{RIW-alt} are the
same, except that in~\eqref{RIW} we use an estimate of the inverse
variance that is not necessarily given by the negative observed
Fisher information.

%s3 ###
\section{Generating a Complete Data Set}
\label{gen-cd}
Let $S(t  | Z)$ be the survival function for an individual with
covariate vector $Z$.  The proportional hazards model may be
reformulated as
%
%e7 ###
\begin{equation}
  \label{cox-model-3}
  S(t |  Z) = (S_0(t))^{\exp(\beta' Z)},
\end{equation}
where $S_0$ is the survival function for an individual with
covariate vector $0$.  Models~\eqref{cox-model-2}
and~\eqref{cox-model-3} are equivalent in the continuous case, for
which the survival function and corresponding cumulative hazard
function are related via $S(t) = \exp(-\Lambda(t))$.  In
general,~\eqref{cox-model-2} and~\eqref{cox-model-3} are not the
same, and it is important to decide on the specification of the Cox
model, and here we take~\eqref{cox-model-3} as our definition.
There are reasons why~\eqref{cox-model-3} is more sensible; see
Kalbfleisch and Prentice (\citeyear{KalbfleischPrentice1980}, Section 4.6).

For an individual with covariate $0$, the survival function and the
cumulative hazard function are related via the product integral
$S_0(t) = \prod_{s \leq t}(1 - \Lambda_0(ds))$
(Gill and Johansen, \citeyear{GillJohansen1990}), so by~\eqref{cox-model-3} the survival
function for an individual with covariate $Z$ is given by
%
%e8 ###
\begin{equation}
  \label{SZ}
  S(t  | Z) = \bigl\{ \textstyle \prod_{s \leq t}\bigl(1 -
  \Lambda_0(ds)\bigr) \bigr\}^{\exp(\beta'Z)}.
\end{equation}
Suppose that the survival time for individual $i$ is censored, that is,
we observe $T_i$ and $Z_i$ and we know that $X_i > T_i$.  We form
$\hat{S}(t | Z_i)$ by substituting $\hat{\Lambda}_0$ and
$\hat{\beta}$ for $\Lambda_0$ and $\beta$ in~\eqref{SZ}, and
generate $X_i$ from this distribution conditional on its being
greater than $T_i$.  We do this for all censored observations, and
the expectations in~\eqref{RI1-den} and~\eqref{vars-ob-co} can be
estimated by Monte Carlo.  Standard software gives $\hat{\Lambda}_0$
and the corresponding $\hat{S}_0$, so this scheme is easy to carry
out.  R functions to implement this scheme and to calculate the
criteria~$\mathcal{R}I_1$ and~$\mathcal{R}I_W$ are available from me upon
request.
\iffalse\fi

\end{document}